\documentclass[
aps,%
12pt,%
final,%
notitlepage,%
oneside,%
onecolumn,%
nobibnotes,%
nofootinbib,%
superscriptaddress,%
noshowpacs,%
centertags]%
{revtex4}

\begin{document}
\selectlanguage{english}

\thispagestyle{empty}

\begin{center}
{\large \bf Where are the Pevatrons that Form the Knee in the Spectrum of the Cosmic Ray Nucleon Component around 4 PeV?}\bigskip

{\bf \copyright{} 2023 A.~A.~Lagutin$^{\pmb{1)}^*}$ and N.~V.~Volkov$^{\pmb{1)}^{**}}$}\blfootnote{$^{1)}$Altai State University, Radiophysics and Theoretical Physics Department}\blfootnote{\;$^*$E-mail: {\tt lagutin@theory.asu.ru}}\blfootnote{\;$^{**}$E-mail: {\tt volkov@theory.asu.ru}}

\begin{quotation}
\noindent{\bf Abstract---}The paper discusses an approach that made it possible to estimate the distance to the nearest pevatrons, which form a knee in the spectrum of the cosmic ray nucleon component of about $4$~PeV. It is based on the spectra of nucleons and electrons obtained by the authors in the framework of the superdiffusion model of nonclassical cosmic rays diffusion, which have a knee, on the assumption that nucleons and electrons are accelerated by the same type sources and their propagation in an inhomogeneous turbulent galactic medium is characterized by the same diffusion coefficient, and also on the knee in the spectrum of the electronic component in the region of $0.9$~TeV, established in the DAMPE experiment.

It is shown that pevatrons, which form a knee in the spectrum of the cosmic ray nucleon component of about
$4$~PeV, are located at distances of the order of $0.75$~kpc from the Earth.
\end{quotation}
\end{center}

\section{Introduction}

Despite more than 100 years of research, the spatial distribution of the main cosmic rays (CRs) sources and the mechanisms of particle acceleration in them have not been finally established. To solve the problem of searching of galactic CR sources the energies of the order of $10^{15}$~eV is a key place because at these energies the CR spectrum has a break (so-called ``knee''). Today it is generally accepted that CRs with energies around the knee are mainly of galactic origin, and their sources are called pevatrons.

The search of galactic pevatrons is currently one of the priority task solved jointly by all ground-based and orbital astrophysical observatories operating in the very high energy region (see for example reviews~\cite{HAWC:2020,Cristofari:2021,Fermi:2022,LHAASO:2023}). One of the most important results achieved is the detection of gamma rays with energies above 100~TeV, clearly indicating that there is an effective acceleration of CR particles up to energies of the order of $10^{15}$~eV. For many years, the scientific community has been dominated by the hypothesis that the main sources capable of accelerating CRs to such energies are supernova remnants (SNRs). Despite this, today there are no reliable experimental data confirming the fact that supernovae accelerate CR nuclei to energies of $\sim 4$~PeV, i.e., to the knee energy in the CRs spectrum~\cite{Caprioli:2011,Blasi:2019,Evoli:2021,Vieu:2023,Cardillo:2023}. The detection of ultrahigh-energy gamma-rays from regions not associated with SNRs indicates that there are other astrophysical objects that can claim the role of pevatrons (see materials of HONEST Workshops~\cite{HONEST:2022}).

Main goal of this paper is to discuss an approach that made it possible to estimate the distance to the nearest pevatrons, which form a knee in the spectrum of the CRs nuclear component about $4$~PeV.

\section{Proposed approach}

The key elements of the proposed approach are based on the following results and assumptions.

\begin{itemize}
\item Our approach is based on the spectra of nuclei and electrons obtained by the authors in the framework of the superdiffusion model of nonclassical CRs diffusion, which have a knee. The main provisions of the nonclassical diffusion model and the results of its application for the interpretation of the spectra of the leptonic and nuclear components of CRs, as well as the spectrum and mass composition in the ultrahigh energy region, are given in our previous papers~\cite{Lagutin:2001,Lagutin:2003,Lagutin:2004a,Lagutin:2004b,Lagutin:2009,Lagutin:2021a,Lagutin:2021b,Lagutin:2023}.
\item We assume that nuclei and high energy electrons and positrons are accelerated by the same type sources and their propagation in an inhomogeneous (fractal-like) turbulent galactic medium is characterized by the same diffusion coefficient.
\item We use the fact that there is a knee in the high-energy cosmic-ray electrons plus positrons spectrum in the region $\sim 0.9$~TeV. Early indications of the presence of this knee were obtained by ground-based Cherenkov detectors of the H.E.S.S. collaboration~\cite{HESS:2008}. Recent results of direct measurements by DAMPE~\cite{DAMPE:2017} and CALET~\cite{CALET:2017} space observatories confirmed the presence of this knee in the total spectrum of electrons and positrons.
\end{itemize}

\section{The nonclassical CRs diffusion equations}

For the first time, the equations of superdiffusion of cosmic rays without taking into account energy losses and taking them into account were proposed in our works~\cite{Lagutin:2001,Lagutin:2004a}.

For the density of particles $N(\rr,t,E)$ with energy $E$ at the location $\rr$ and time $t$, generated
in a fractal-like medium by Galactic sources with a distribution density $S(\rr,t,E)$, is written as
\begin{equation}\label{eq:superdiffnuc}
\frac{\partial N(\rr,t,E)}{\partial t}= -D(E,\alpha)(-\Delta)^{\alpha/2} N(\rr,t,E) + S(\rr,t,E),
\end{equation}


\begin{equation}\label{eq:superdiffelpos}
\frac{\partial N(\rr,t,E)}{\partial t}= -D(E,\alpha)(-\Delta)^{\alpha/2} N(\rr,t,E)+ \dfrac{\partial B(E)N(\rr,t,E)}{\partial E} + S(\rr,t,E).
\end{equation}
In these equation $D(E,\alpha)=D_0(\alpha)E^{\delta}$ is the anomalous diffusivity; $(-\Delta)^{\alpha/2}$ is the fractional Laplacian~\cite{Samko:1993} (``Riesz operator'') (reflects a nonlocality of the diffusion process of particles in the interstellar medium); $B(E)$ is the mean rate of continuous energy losses of electrons and positrons.


It should be noted that when $\alpha=2$ from Eqs.~\eqref{eq:superdiffnuc} and~\eqref{eq:superdiffelpos} we obtain the normal diffusion Ginzburg-Syrovatskii equations.

In calculations of the spectrum of electrons and positrons, the main mechanisms of energy lossess are taken into account, i.e. ionization, bremsstrahlung, synchrotron and inverse Compton losses. In a recent paper~\cite{Fang:2021} it was shown that in the relativistic regime (Klein-Nishina mode) the threshold energy of inverse Compton scattering is reached even when electrons interact with photons of visible radiation. The cross sections for the interaction of electrons with background photons become much smaller than the Thomson cross section traditionally used in calculations. As a result, the rate of energy lossess of electrons during interaction with the background electromagnetic radiation of the Galaxy decreases (the Klein-Nishina effect). In this paper, in calculating the spectra of electrons and positrons with energies $E > 100$~GeV, the approximation expressions proposed in~\cite{Fang:2021} were used to take into account the Klein-Nishina effect.

The solution of the superdiffusion Eqs.~\eqref{eq:superdiffnuc} and~\eqref{eq:superdiffelpos} was found by the Green's function method in both cases with energy losses and without ones. To solve equation~\eqref{eq:superdiffelpos} we use the Syrovatskii functions~\cite{Ginzburg:1964}

\begin{equation}\label{eq:lambdatau}
\lambda(E,E_0)=\int\limits_E^{E_0}\frac{D(E')}{B(E')}dE',\qquad \tau(E,E_0)=\int\limits_E^{E_0}\frac{dE'}{B(E')}.
\end{equation}

\begin{figure}[ht!]
\setcaptionmargin{5mm}
\onelinecaptionsfalse 
\includegraphics[width=.8\textwidth]{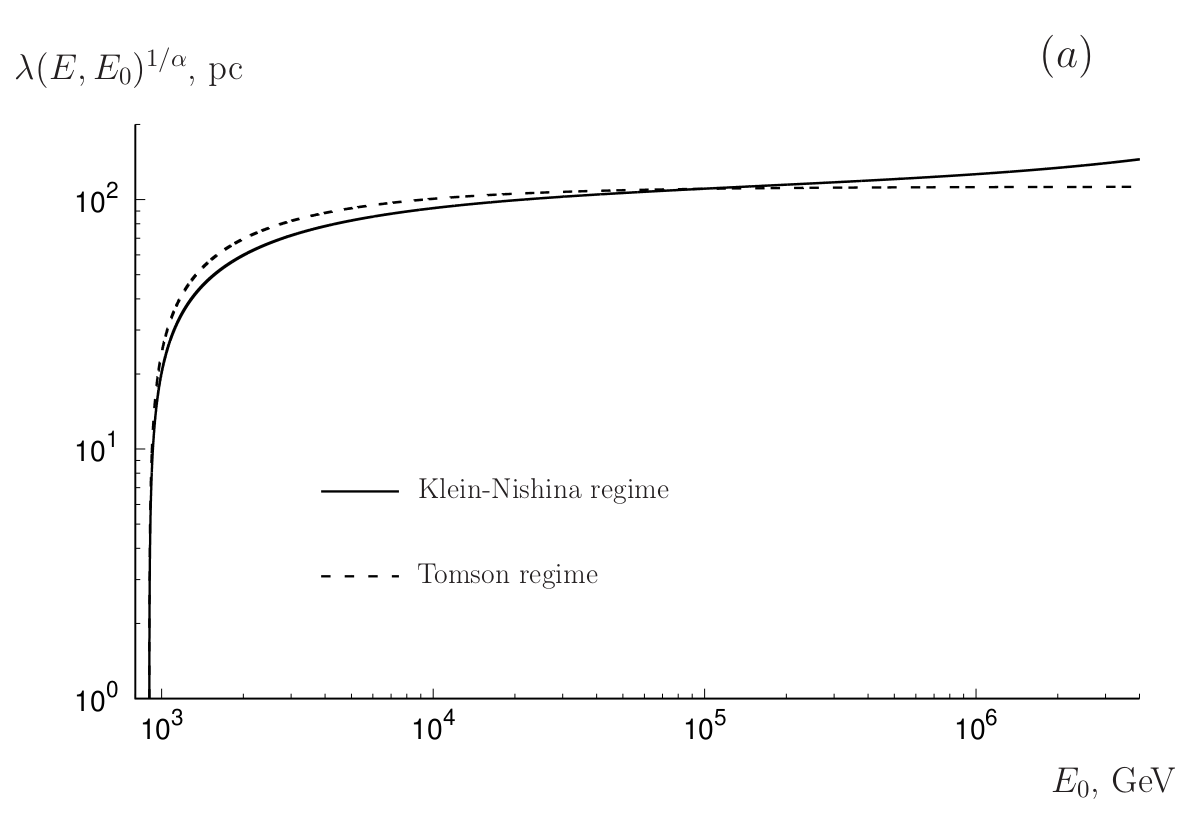}
\includegraphics[width=.8\textwidth]{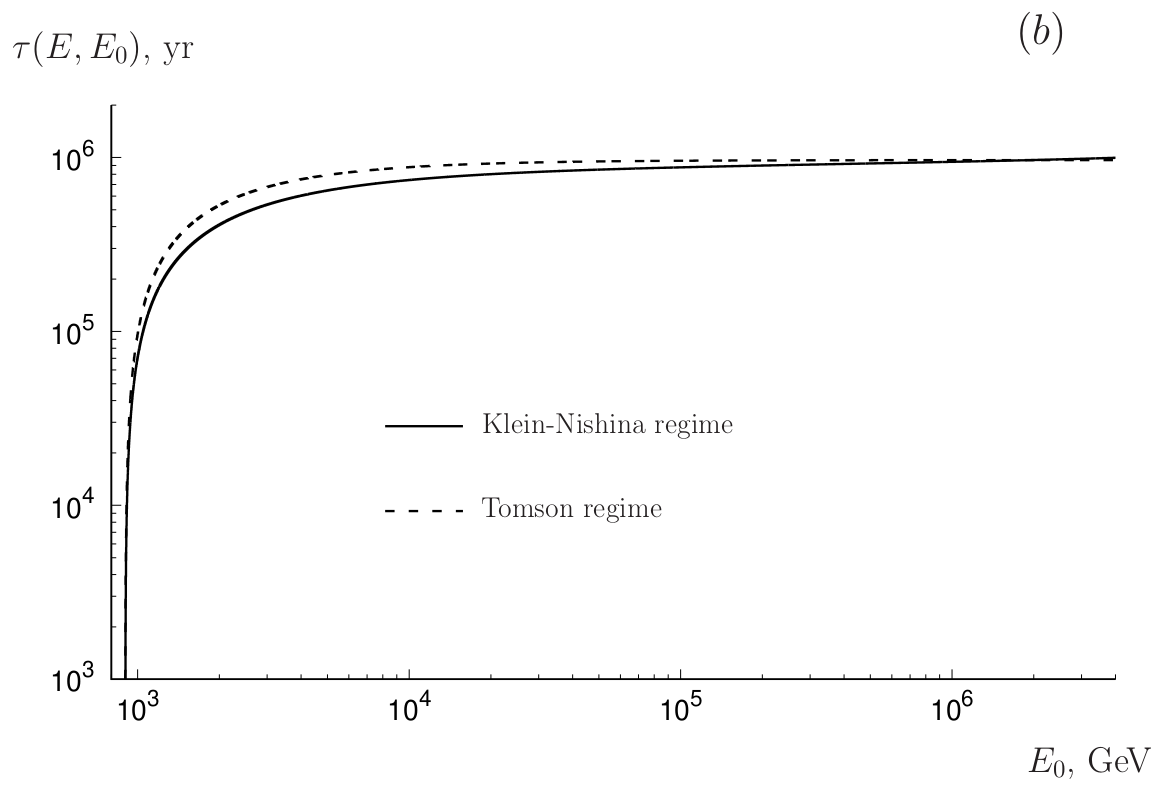}
\caption{Dependences of the Syrovatskii functions $\lambda(E,E_0)^{1/\alpha}$ and $\tau(E,E_0)$ on the initial energy $E_0$ of the particles for the different regimes of energy losses}\label{fig:fig1}
\end{figure}

The $\lambda(E,E_0)^{1/\alpha}$ function should be understood as the average distance over which the particle has diffused, taking into account the energy losses from the initial value $E_0$ to the energy $E$. The $\tau(E,E_0)$ function gives the ``cooling time'' from $E_0$ energy to $E$. Fig.~\ref{fig:fig1} show the behavior of the functions  $\lambda(E,E_0)^{1/\alpha}$ and $\tau(E,E_0)$ for different values of the initial energy $E_0$ of the particles. The calculations were carried out for the both Klein-Nishina and Thomson regimes of energy losses. It can be seen from the Fig.~\ref{fig:fig1}(a) that if the particle energy at the point of observation is equal to $0.9$~TeV, then for any mode of energy lossess from the initial value of $4$~PeV, the diffusion radius of such particles is about $100-200$~pc. It can be seen from Fig. \ref{fig:fig1}(b), the ``cooling time'' of such particles varies from $10^5$ to $10^6$ years.

For point instant source $S(\rr,t,E) = S_0E^{-p} \delta(\rr)\delta(t)$ solutions of the superdiffusion Eqs.~\eqref{eq:superdiffnuc} and~\eqref{eq:superdiffelpos} take the form:

\begin{itemize}
\item for the nuclear component of CR (without energy losses)
\begin{equation}~\label{eq:superdiffnucsol}
N(\rr,t,E) = S_0E^{-p} (D(E,\alpha)t)^{-3/\alpha} g_3^{(\alpha)}(|\rr|(D(E,\alpha)t)^{-1/\alpha}).
\end{equation}

\item for the leptonic component of CR (with energy losses)
\begin{equation}~\label{eq:superdiffelpossol}
N(\rr,t,E) = \frac{S_0E^{-p}}{B(E)} \lambda(E,E_0)^{-3/\alpha} g_3^{(\alpha)}(|\rr|\lambda(E,E_0)^{-1/\alpha}),
\end{equation}
\end{itemize}

In the expressions~\eqref{eq:superdiffnucsol} and~\eqref{eq:superdiffelpossol} $g_3^{(\alpha)}(r)$ is the probability density of three-dimentional sphericaly-symmetrical stable distribution~\cite{Uchaikin:1999a,Zolotarev:1999}. The key feature of this function is the presence of a knee. In case $\alpha=2$ $g_3^{(\alpha)}(r)$ is the normal distribution.

In the particular case (Tomson regime) for energy losses in the form $B(E) = b E^2$~GeV/s, where $b = 1.1\cdot 10^{-16}$ (GeV s)$^{-1}$ we find the solution of superdiffusion Eq.~\eqref{eq:superdiffelpos}
\begin{equation}~\label{eq:superdiffelpossol2}
N(\rr,t,E) = S_0E^{-p}(1-btE)^{p-2} \lambda(t,E)^{-3/\alpha} g_3^{(\alpha)}(|\rr|\lambda(t,E)^{-1/\alpha}).
\end{equation}
Here
\begin{equation}~\label{eq:lambda}
\lambda(t,E) = D_0(\alpha)E^{\delta} \hat{\lambda}(t,E),\qquad \hat{\lambda}(t,E) = \dfrac{1-(1-btE)^{1-\delta}}{b(1-\delta)E}.
\end{equation}

\begin{figure}[ht!]
\setcaptionmargin{5mm}
\onelinecaptionsfalse 
\includegraphics[width=\textwidth]{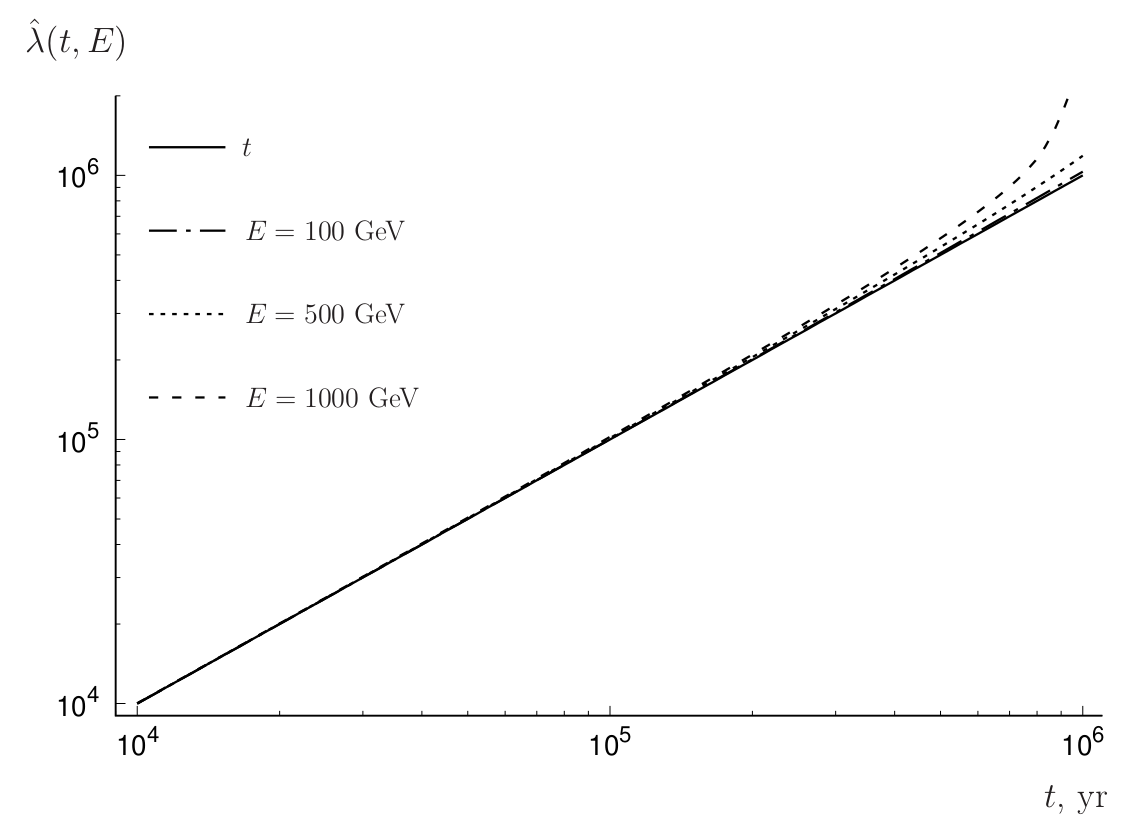}
\caption{Dependence of the function $\hat{\lambda}(t,E)$ on time $t$ for different values of energy $E$}\label{fig:fig2}
\end{figure}

The figure~\ref{fig:fig2} shows the results of calculating the function $\hat{\lambda}(t,E)$ for various values of energy $E$. It should be noted that in wide range of parameters $E$ and $t$ we obtain $\hat{\lambda}(t,E)\equiv t$, i.e. $\lambda(t,E) = D(E,\alpha) t$. Calculations in the Klein-Nishina mode lead to the same conclusion.

\section{Knee in the energy spectrum}

To analyze the energy dependence of the electron concentration, we write solution~\eqref{eq:superdiffelpossol2} of the superdiffusion Eq.~\eqref{eq:superdiffelpos} in the form $N = N_0E^{-\eta}$. It follows from this representation that
$$\eta = -\dfrac{E}{N}\dfrac{\partial N}{\partial E}.$$

Taking into account the property of the stable law~\cite{Uchaikin:1999a} $\dfrac{d g_3^{(\alpha)}(r)}{dr} = -2\pi r g_5^{(\alpha)}(r)$, we find
\begin{equation}~\label{eq:eta}
\eta = 2p - 2 + \dfrac{\delta - 1}{\alpha}\left[3-\dfrac{2\pi r^2}{\lambda(t,E)^{1/\alpha}}\dfrac{g_5^{(\alpha)}(|\rr| \lambda(t,E)^{-1/\alpha})}{g_3^{(\alpha)}(|\rr| \lambda(t,E)^{-1/\alpha})}\right] = 2p - 2 + \dfrac{\delta - 1}{\alpha}\Xi.
\end{equation}
In Eq.~\eqref{eq:eta} $g_5^{(\alpha)}(r)$ is the probability density of five-dimentional stable distribution~\cite{Uchaikin:1999a}.

\begin{figure}[ht!]
\setcaptionmargin{5mm}
\onelinecaptionsfalse 
\includegraphics[width=\textwidth]{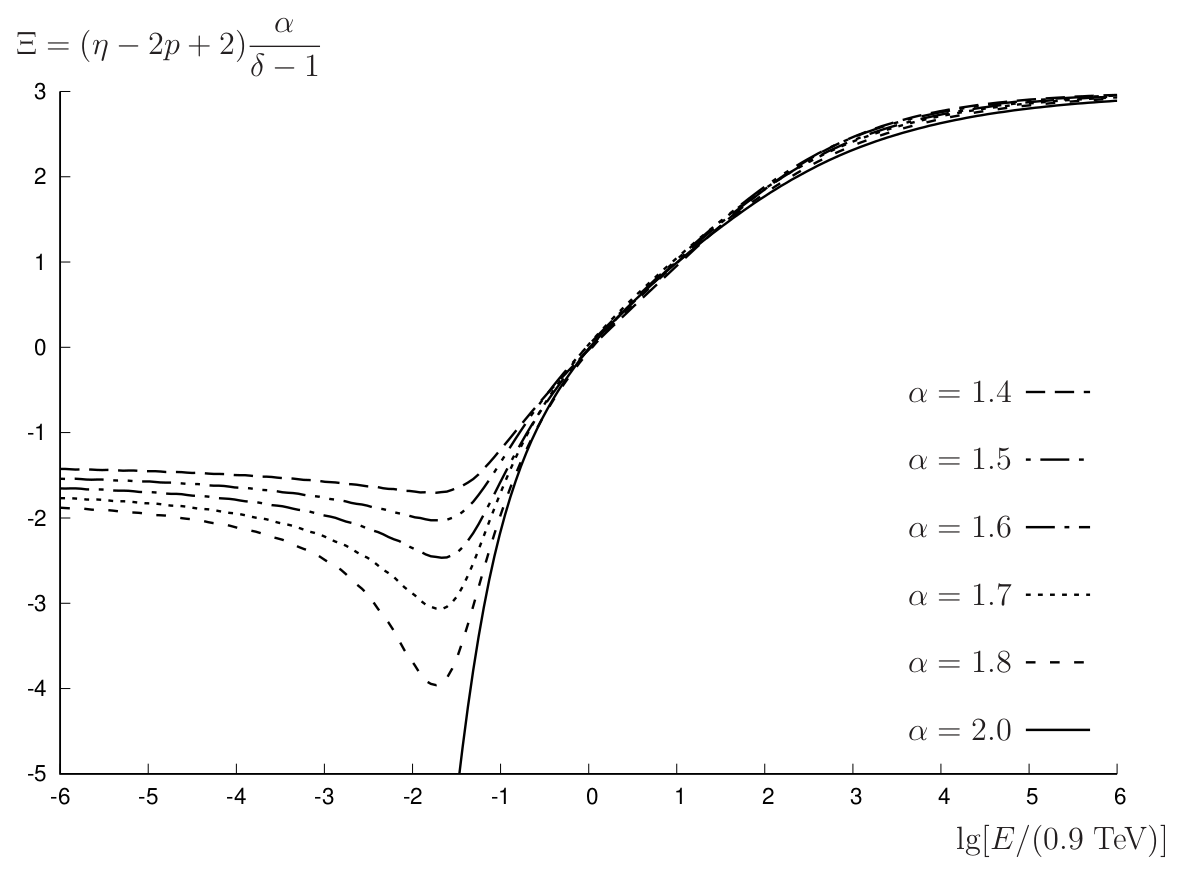}
\caption{Change in the spectral index of the observed electrons on the energy in the case of a point instant source for the different diffusion regimes. $r = 200$~pc}\label{fig:fig3}
\end{figure}

Fig.~\ref{fig:fig3} shows the energy dependence of the index $\Xi$ of the electrons and positrons observed spectrum for various diffusion modes (parameter $\alpha$). It can be seen that the spectrum has a knee in the superdiffusion regime $\alpha < 2$ (exponent $\Xi$ at the knee point of $0.9$ TeV is equal to zero). At the same time, in the normal diffusion regime, the spectral index $\Xi$ increases monotonically with increasing energy. In this case the spectrum has no knee.

It should be noted that the break in the electron spectrum obtained in the framework of the nonclassical diffusion model, as well as comparison with experimental data, will be considered in our next work.

\section{Approach to estimate the distance to the nearest pevatrons}

It was shown in~\cite{Lagutin:2001} that the knee may be due to anomalous CR diffusion in a turbulent (fractal type) interstellar medium, but not the presence of maximum cosmic ray energy in sources. Recently published data of the LHAASO Collaboration on the spectrum of diffuse gamma radiation from the disk of the Galaxy~\cite{LHAASO:2023aug}, which is described by a power-law function with an exponent $-2.99$ in the entire energy region of 10~TeV - 1~PeV may be considered as an indication on the validity of the hypothesis that the knee in the $1-4$~PeV region is  most likely due to transport.

In the framework of nonclassical CRs diffusion the knee is due to the presence of a break in the stable distribution $g_3^{(\alpha)}(r)$ at the value of the argument $r \approx 2.2$. On this basis the expressions~\eqref{eq:superdiffelpossol} and~\eqref{eq:superdiffnucsol} for the spectra obtained in the nonclassical diffusion model make it possible to establish a relationship between the characteristics $(\rr, t, E)$ of the knee points of the nuclear ($n$) and leptonic ($e$) components.

For $n$ component
$$N(\rr,t,E) \sim g_3^{(\alpha)}\left(|\rr| (D(E,\alpha) t)^{-1/\alpha}\right)\Rightarrow r_n (D_0(\alpha) E_n^{\delta} t_n)^{-1/\alpha} = 2.2.$$

For $e$ component
$$N(\rr,t,E) \sim g_3^{(\alpha)}\left(|\rr| \lambda(t,E)^{-1/\alpha}\right)\Rightarrow r_e (D_0(\alpha) E_e^{\delta} \hat{\lambda}(t_e,E_e))^{-1/\alpha} = 2.2.$$

We assume that CR nuclei and high-energy electrons are accelerated by the same types of sources (pevatrons) and their propagation in an inhomogeneous turbulent galactic medium is characterized by the same diffusion coefficient $D_0(\alpha)$. Due to this assumption
\begin{equation}\label{eq:params}
r_n (D_0(\alpha) E_n^{\delta} t_n)^{-1/\alpha} = r_e (D_0(\alpha) E_e^{\delta} \hat{\lambda}(t_e,E_e))^{-1/\alpha}.
\end{equation}

It follows from the Eq.~\eqref{eq:params} that
\begin{equation}\label{eq:rnre}
r_n = r_e \left[\left(\dfrac{E_n}{E_e}\right)^{\delta}\dfrac{t_n}{\hat{\lambda}(t_e,E_e)}\right]^{1/\alpha}\equiv r_e \xi.
\end{equation}

It should be noted that the estimates obtained within the framework of the proposed approach are almost diffusion model independent.

\section{Results}

Parameters of the nonclassical diffusion model and the technology for their self-consistent determination
using the available experimental data on galactic CRs were discussed in our previous papers~\cite{Lagutin:2021b,Lagutin:2023}. The main parameters of the model are given in the Table~\ref{tab:adparams}. Spatiotemporal characteristics nearest galactic CR sources, which include the Geminga, Monogem, and Vela pulsars, are given in~\cite{Lagutin:2009}.

\begin{center}
\begin{table}[htb!]
\centering
\caption{The nonclassical diffusion model parameters}\label{tab:adparams}
\begin{tabular}{|l|l|}
\hline
\textbf{Parameter} & \textbf{Value} \\
\hline
$p$ & $2.85$ \\
\hline
$\delta$ & $0.27$ \\
\hline
$D_0(\alpha)$ & $10^{-3}$~pc$^{1.7}$yr$^{-1}$\\
\hline
$\alpha$ & $1.7$\\
\hline
$t_e$, $t_n$ & $10^5$~yr \\
\hline
$E_e$ & $0.9$ TeV\\
\hline
$E_n$ & $0.65$ PeV\\
\hline
\end{tabular}
\end{table}

\begin{table}[htb!]
\caption{Space-time parameters of the most likely candidates for pevatrons according to~\cite{Kobayashi:2004,Lozinskaya:1992}}\label{tab:pevparams}
\begin{tabular}{l|l|l}
\hline
Source & $r$, pc & $t, 10^5$~yr\\
\hline
Monoceros & 600 & 0.46 \\
Cyg. Loop & 770 & 0.20 \\
CTB 13 & 600 & 0.32 \\
S 149 & 700 & 0.43 \\
STB 72 & 700 & 0.32 \\
CTB 1 & 900 & 0.47 \\
HB 21 & 800 & 0.23 \\
HB 9 & 800 & 0.27 \\
\hline
\end{tabular}
\end{table}
\end{center}

The ``lifetime'' of the CR electrons is described by the function $\tau(E,E_0)$ from Eq.~\eqref{eq:lambdatau}.
It follows from this that the TeV-energy electrons observed on Earth were produced by sources $\sim 10^5$~years ago. During this time, in the superdiffusion mode $\overline{r^2}\sim 2D(E,\alpha)t^{3-\alpha}$~\cite{Perri:2015}, diffusion radius of electrons $r_e\sim 200$~pc.

To obtain estimates of the distance to the nearest pevatrons, we use the value of the knee energy of the proton spectrum $E_n = 650$~TeV that has been obtained in~\cite{Bartoli:2015,Lagutin:2017}. From Eq.~\eqref{eq:rnre} we obtain that $r_n = 3.75 r_e$. Thus pevatrons, which form a knee in the spectrum of the nuclear component of CRs of about $4$~PeV, are located at distances of the order of $0.75$~kpc from the Earth.

The most likely candidates for pevatrons are shown in the Table~\ref{tab:pevparams}.

\section{Conclusions}

An approach that makes it possible to estimate the distance to the nearest pevatrons, which form a knee in the spectrum of the CRs nuclear component of about $4$~PeV, has been formulated.

It is based on the CRs spectra of nuclei and leptons obtained by the authors in the framework of the superdiffusion model of nonclassical diffusion, which have a knee, on the assumption that nuclei and electrons are accelerated by the same types of sources and their propagation in an inhomogeneous turbulent galactic medium is characterized by the same diffusion coefficient and also on the knee in the spectrum of the leptonic component in the region of $0.9$~TeV, established in the DAMPE and CALET experiments.

It has been established that pevatrons, which form a knee in the spectrum of the CRs nucleon component of about $4$~PeV, are located at distances of the order of $0.75$~kpc from the Earth.

\begin{acknowledgments}
The work is supported by the the Russian Science Foundation (grant no. 23-72-00057).
\end{acknowledgments}

\end{document}